\documentclass[aps,prl,reprint,twocolumn,superscriptaddress]{revtex4-1}

\usepackage{euscript}
\usepackage{amssymb}
\usepackage{amsfonts}
\usepackage{amsbsy}
\usepackage{epsfig}
\usepackage{amsthm}
\usepackage{amscd}
\usepackage{amstext}
\usepackage{verbatim}
\usepackage{amsmath}
\usepackage{cancel}
\usepackage{capt-of}
\usepackage{empheq}
\usepackage{subfigure}
\usepackage{xcolor}

\usepackage[pdftex]{hyperref}
\hypersetup{ 
colorlinks=true, 
linkcolor=black, 
citecolor=red, 
}

\renewcommand{\Im}[0]{\operatorname{Im}}

\usepackage{amssymb}

\def\ben{\begin{equation}}
\def\een{\end{equation}}

\let\a=\alpha \let\b=\beta

\let\w=\omega    
   
 \let\W=\Omega

\let\pa=\partial
\def\be{\begin{equation}}
\def\ee{\end{equation}}
\def\beq{\begin{equation}}
\def\eeq{\end{equation}}
\def\ba{\begin{array}}
\def\ea{\end{array}}
\def\del{\partial}

\def\wtd{\widetilde}

\def\dalemb#1#2{{\vbox{\hrule height .#2pt
       \hbox{\vrule width.#2pt height#1pt \kern#1pt
               \vrule width.#2pt}
       \hrule height.#2pt}}}

\newcommand{\bea}{\begin{eqnarray}}
\newcommand{\eea}{\end{eqnarray}}
\newcommand{\tr}{{\rm tr} }

\makeatletter
\newcommand*\bigcdot{\mathpalette\bigcdot@{.5}}
\newcommand*\bigcdot@[2]{\mathbin{\vcenter{\hbox{\scalebox{#2}{$\m@th#1\bullet$}}}}}
\makeatother

\renewcommand{\eqref}[1]{(\ref{#1})}

\def\Im{{{\frak{Im}}}}

\def\Lag{{\mathcal{L}}}

\def\Ts{{\mathcal{T}}}

\def\ocal{{\mathcal{O}}}

\def\bk{\boldsymbol{k}}
\def\bx{\boldsymbol{x}}

\renewcommand{\Im}[0]{\operatorname{Im}}

\begin{document}
\frenchspacing

\title{Locality Bound for Dissipative Quantum Transport}
\author{Xizhi Han and Sean A. Hartnoll \\ {\it Department of Physics, Stanford University,}\\
{\it Stanford, California, USA}
}

\date{}


\begin{abstract}
We prove an upper bound on the diffusivity of a dissipative, local and translation invariant quantum Markovian spin system: $D \leq D_0 + \left(\a \, v_\text{LR} \tau + \beta \, \xi \right) v_\text{C}$. Here $v_\text{LR}$ is the Lieb-Robinson velocity, $ v_\text{C}$ is a velocity defined by the current operator, $\tau$ is the decoherence time, $\xi$ is the range of interactions, $D_0$ is a decoherence-induced microscopic diffusivity and $\a$ and $\b$ are precisely defined dimensionless coefficients. The bound constrains quantum transport by quantities that can either be obtained from the microscopic interactions ($D_0, v_\text{LR}, v_\text{C},\xi$) or else determined from independent local non-transport measurements ($\tau,\a,\beta$). We illustrate the general result with the case of a spin half XXZ chain with on-site dephasing. Our result generalizes the Lieb-Robinson bound to constrain the sub-ballistic diffusion of conserved densities in a dissipative setting.
\end{abstract} 

\maketitle

{\it Introduction.---} Quantum transport processes are at the heart of experimental studies of unconventional metals \cite{RevModPhys.75.1085, hussey, Bruin804}, ultracold atomic gases \cite{schneider2012fermionic, PhysRevLett.108.093002, koschorreck2013universal, hild14, chien2015quantum,  PhysRevLett.118.130405} and potential spintronic systems \cite{Wolf1488, RevModPhys.76.323, awschalom2007challenges, bogani2008molecular, han2014graphene}. It is crucial to have theoretical tools that connect transport observables to microscopic processes. In quasiparticle systems such as conventional metals, Fermi liquid theory and Boltzmann equations offer an excellent and well-understood handle on transport \cite{ziman}. For strongly quantum transport regimes, however, there are many fewer tools available. Controlled theoretical work with realistic interactions has largely been restricted to numerics in one spatial dimension  \cite{SCHOLLWOCK2011, kar12, Leviatan:2017, knap17, zal18}.

For general ballistic systems, several important, rigorous bounds on quantum transport have been established. The Mazur-Suzuki inequality bounds the Drude weight in terms of the overlap of currents with conserved charges \cite{MAZUR69, SUZUKI71, pro11}. The Lieb-Robinson velocity $v_\text{LR}$ bounds the propagation of linearly dispersing collective modes, such as spin waves
\cite{lieb1972, PhysRevLett.104.190401}.

Many important quantum transport processes are diffusive rather than ballistic \cite{hartnoll15}. A lower bound on the high-temperature diffusivity has been established for certain systems with integrability or additional symmetries \cite{pro14, PhysRevLett.119.080602}. Recently, it was argued that in general local systems, Lieb-Robinson causality requires that the diffusivity be upper bounded as $D \lesssim v_\text{LR}^2 \tau_\text{th}$ \cite{Hartman:2017hhp}. Here $\tau_\text{th}$ is a `local thermalization time'. This relation usefully identifies key physical ingredients that constrain diffusive transport. However, it is not totally satisfactory because a numerical prefactor is undetermined and furthermore the timescale $\tau_\text{th}$ was not precisely defined.

In this work we prove a rigorous and precise upper bound on the diffusivity of dissipative quantum Markovian spin systems. The full result is given in (\ref{eq:fullbound}) below. In the limit of long decoherence time $\tau$, the bound takes the form $D \leq \alpha v_\text{LR} v_\text{C} \tau$. This expression is the dissipative counterpart of the earlier bound \cite{Hartman:2017hhp}, and all quantities on the right hand side will now be precisely defined. The velocities $v_\text{LR}$ and $v_\text{C}$ are straightforwardly computed given a microscopic Hamiltonian while the dimensionless coefficient $\a$ and decoherence time $\tau$ can be independently and unambiguously determined from local non-transport observables. Therefore, this bound can be precisely verified in experiments. It generalizes the Lieb-Robinson bound to the diffusive behavior of conserved densities, in the context of dissipative quantum Markovian dynamics.

{\it Translation invariant Lindbladian dynamics.---} Non-unitary quantum dynamics describes the quantum evolution of a dissipative system coupled to an external environment. On timescales much longer than the relaxation time of the reservoir, the dynamics can be well approximated as Markovian and hence described by the Lindblad equation \cite{lindblad1976, bre07}. The final state of Lindbladian non-unitary dynamics is expected to be an infinite temperature generalized Gibbs ensemble, so our diffusive dynamics occurs close to this state.

We assume that the external bath couples locally in space to the degrees of freedom of interest, and preserves spatial translation invariance. 
In this case, the most general Heisenberg equation of motion for an operator $O(t)$ on a lattice takes the Lindblad form
\be
\dot{O}  = i \sum_{\bx} [H_{\bx}, O] + c \sum_{\bx, \a} \left( 2 L_{\bx}^{\a \dagger} O L_{\bx}^{\a} -  \left\{ L_{\bx}^{\a \dagger} L_{\bx}^{\a}, O\right\} \right), \label{eq:1}
\ee
where $\bx$ is the lattice index and $H_{\bx}$ is a term in the Hamiltonian localized near lattice site $\bx$. The anticommutator $\{A,B\} = AB +BA$. The $L_{\bx}^\alpha$ are decoherence operators localized near site $\bx$ and $c \geq 0$ is the decoherence strength.
It will be important that a Lieb-Robinson velocity exists for such local Lindbladian dynamics \cite{PhysRevLett.104.190401, nachtergaele2011lieb}.

Throughout, we illustrate our general formalism and results with the example of an infinite, spin-half antiferromagnetic XXZ chain with on-site dephasing:
\begin{align}\label{eq:model}
    H_x  = X_x X_{x + 1} + Y_x Y_{x + 1} + \Delta Z_x Z_{x + 1} \,, \quad L_x  = Z_x \,.
\end{align}
Here $X_x, Y_x, Z_x$ are Pauli matrices acting on spin $x \in \mathbb{Z}$ and $\Delta > 0$ is the anisotropy. The dephasing Lindbladian is a common phenomenological description of decoherence due to coupling to a photon or phonon bath \cite{RevModPhys.70.101}. Diffusion in this model was studied numerically in \cite{z, PhysRevLett.106.220601, men13, pro16, ljubotina2017spin}, and we will compare with those results. Our approach, however, is not limited to one dimensional models.

The model (\ref{eq:model}) conserves spin: $\sum_{x} \dot{Z}_x = 0$. More generally, we require a local charge operator $C$ such that
\be\label{eq:conserved}
\sum_{\bx} \dot{C}_{\bx} = 0,
\ee
where $C_{\bx}$ is the operator $C$ translated to site $\bx$. A conserved operator in the sense of (\ref{eq:conserved}) has important consequences for the dynamics on the longest timescales, after all non-conserved operators have decayed. A single, scalar conserved operator is expected to lead to a diffusive mode with long wavelength dispersion
$\omega(\bk) = - i D k^2 + \ldots$, see e.g. \cite{cha95}.  Here $D$ is the diffusivity and $\ldots$ denotes terms of higher order in the wavevector $\bk$. Our objective in the remainder is to connect the microscopic Lindbladian dynamics (\ref{eq:1}) to the long wavelength hydrodynamic mode, and in this way bound the diffusivity $D$ in terms of microscopic quantities.

To exploit the translation invariance of the dynamics, we introduce the linear space of operators $\ocal_{\bk}$ with wavevector $\bk$, defined to be the space of all operators $O$ on an infinite lattice $\Lambda$ such that
\be
\Ts_{\bx}[O] = O\, e^{i \bk \cdot \bx},
\ee
where $\Ts_{\bx}$ translates operators by a vector $\bx$. It will be useful to take the following basis of operators in $\ocal_{\bk}$. Fix an origin of the lattice and a direction $\hat{\bk}$ of the wavevector. We can then write the basis elements of $\ocal_{\bk}$ as
\be
    |O_a) \equiv (O_a)_{\bk} \equiv \sum_{\bx} \Ts_{\bx}[O_a] e^{-i \bk \cdot \bx}, \label{eq:7}
\ee
where $\{O_a\}$ is the set of product operators that are localized in the region $\{\bx \in \Lambda \,|\, \bx \cdot \hat{\bk} \geq 0\}$ and are not the identity at the origin \footnote{Also, the identity operator $I$ itself doesn't contribute to $\ocal_{\bk}$ for ${\bk}\neq 0$ because then $(I)_{\bk} = 0$ in (\ref{eq:7})\label{foot:id}}. We drop the $\bk$ label on the $|O_a)$ to avoid clutter, this basis gives a canonical isomorphism between the different $\ocal_{\bk}$.  For the example of the XXZ chain, the $\{O_a\}$ are strings of Pauli operators starting at the origin:
$X_0, Y_0, Z_0, X_0 X_1, X_0 Y_1, \ldots, X_0 I_1 Y_2, \ldots,$
where subscripts are lattice indices $x \in \mathbb{Z}$. The corresponding basis elements in (\ref{eq:7}) are then operators such as
$(X_0 Y_1)_k = \ldots + X_{-1} Y_0 e^{i k} +  X_0 Y_1 + X_1 Y_2 e^{-i k} + X_2 Y_3 e^{-2 i k} + \ldots$,
from which it is clear that $(X_0 Y_1)_k$ and $(X_1 Y_2)_k$ only differ by a phase prefactor. This is why the operators must be taken to start at $x = 0$.

Translational symmetry implies that the $\ocal_{\bk}$ are preserved by time evolution. Therefore, it is possible to diagonalize $\del_t$ in each $\bk$-sector. An eigenoperator $O_{\bk} \in \ocal_{\bk}$ satisfies
\be\label{eq:oodot}
\dot{O}_{\bk} = - i \w(O, {\bk}) O_{\bk},
\ee
for some $\w(O, \bk) \in \mathbb{C}$ and with $\Im{\w} \leq 0$. Note that $i \pa_t$ is not Hermitian but the negative imaginary part of its eigenvalues means that time evolution is stable. Diffusion is then described by a coarse-grained charge operator $\wtd{C}_{\bk}$ that is an eigenoperator of $\del_t$ with
\be
    \w(\wtd{C}, {\bk}) \equiv \W_{\bk} = - i D k^2 + o(k^2) \,, \label{eq:12}
\ee
which defines the diffusivity $D$ of the conserved charge. More generally $D$ may depend on the direction of $\bk$, and this definition works for any fixed direction of $\bk$. We will obtain the operator $\wtd{C}_{\bk}$ explicitly below.

We are able to discuss diffusion as an operator equation, as in (\ref{eq:oodot}) and (\ref{eq:12}) above, because decoherence causes operator norms to decay. This is a significant technical simplification relative to the case of unitary evolution at finite temperature, where diffusion only occurs within thermal expectation values. In the following section we compute $\W_{\bk}$ in small $k$ perturbation theory. This will give an explicit expression for $D$.

{\it Perturbation theory at small wavevector.---}
At small $k$, we can expand $\del_t|_{\ocal_{\bk}}$ in $k$. Fixing a direction of $\bk$:
\be
\del_t|_{\ocal_{\bk}} = \Lag \equiv \sum_{n \geq 0} k^n \Lag^n, \label{eq:14}
\ee
which defines superoperators $\Lag^n$. For example, in the XXZ chain, the operator $(Z_0)_k \in \ocal_k$ obeys
\begin{align}
(\dot{Z}_0)_k 
 = 2 (e^{-i k} - 1) (X_0 Y_1)_k - 2 (e^{-i k} - 1) (Y_0 X_1)_k \,.\label{eq:13}
\end{align}
With respect to the Pauli string basis, $\del_t|_{\ocal_k}$ is represented as a $k$-dependent matrix. Expanding the coefficients of the basis elements in (\ref{eq:13}) at small $k$ we obtain components of the superoperators in (\ref{eq:14}). As expected from conservation of $Z$, $\Lag^0 |Z_0) = 0$, while $\Lag^1 |Z_0) = 2 i |Y_0 X_1) - 2 i |X_0 Y_1)$ and $\Lag^2 |Z_0 ) = |Y_0 X_1) - |X_0 Y_1)$.

The eigenvalue $- i \Omega_{\bk}$ of $\del_t|_{\ocal_{\bk}}$ can be found using 
standard second-order perturbation theory in small $\bk$ (cf. the memory matrix formalism \cite{for75}). At $\bk = 0$ we know that the eigenvector is the conserved charge $|C)$, with vanishing eigenvalue. Therefore, up to order $k^2$:
\begin{align}
     - i \Omega_{\bk} & = 	k (C | \mathcal{L}^1 | C) \label{eq:16} \\
    & + k^2 \left[(C | \mathcal{L}^2 | C) - \sum_{E^0_a \neq 0} (C | \mathcal{L}^1 | E^0_a) \frac{1}{E^0_a} (E^0_a| \mathcal{L}^1 | C)\right].  \nonumber
\end{align}
The basis vectors $|E^0_a)$ are given by linear combinations of the $|O_a)$ in (\ref{eq:7}) with ${\mathcal L}^0 |E^0_a) = E^0_a |E^0_a)$. The corresponding eigenoperator $|\wtd{C}) = |C) - k \sum_{E^0_a \neq 0} | E^0_a) \frac{1}{E_a^0} (E^0_a| \mathcal{L}^1 | C)$ is the dressed charge operator to this order. We are assuming that the only operator with $E_a^0 = 0$ is the single conserved charge $C$. It is straightforward to extend our analysis to a finite number of conserved charges. We will be more precise about the absence of additional slow operators in the following section. The superoperator $\del_t$ is not antihermitian in general and the eigenoperators $|E^0_a)$ are not necessarily orthogonal. The above perturbation theory formulae retain their standard form, but $(E^0_a | O)$ is defined to be the coefficient in front of $|E^0_a)$ in the expansion of $|O )$ in the basis $\{|E^0_a)\}$. The $(E^0_a|$ are elements of a dual vector space to that spanned by the $|E^0_a)$, and hence have opposite dimensionality. In the case that the operators $E^0_a$ are orthogonal, i.e. $\tr (E_a^{0 \dagger} E^0_b) = \delta_{a b} \tr( E_a^{0 \dagger} E^0_a)$ for some given $\bk$, then we can write $(E^0_a | O) = \tr (E_a^{0 \dagger} O) / \tr (E_a^{0 \dagger} E^0_a)$, as usual.

\begin{figure}[!ht]
    \centering
	\includegraphics[width=0.95\columnwidth]{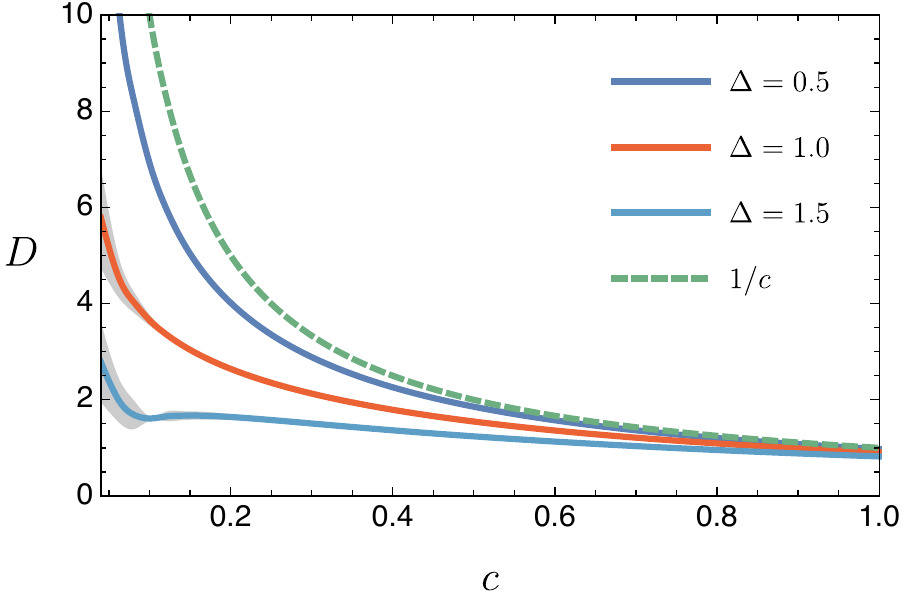}
    \caption{Diffusivity $D$ of the dissipative XXZ model versus dephasing strength $c$, with anisotropies $\Delta = 0.5, 1.0, 1.5$. The asymptotic behavior $D \sim 1 / c$ is also shown. Operator spaces are truncated in numerics so that only Pauli string operators of length at most $n = 7$ are kept. Finite-size effects are strong for small $c$ and indicated by the shaded region, which is estimated from truncations with $n = 6, 8$. \label{fig1}}
\end{figure}

Our main objective is to use the expression (\ref{eq:16}) to bound the diffusivity (\ref{eq:12}) in generality. However, in simple models such as the dissipative XXZ chain (\ref{eq:model}) it is possible to compute the diffusivity by evaluating (\ref{eq:16}). The on-site dephasing in that model suppresses Pauli strings with $X$ and $Y$ terms. For example: $\mathcal{L}_{\mathrm{dis}}[X_0] = - 4 X_0$ and  $\mathcal{L}_{\mathrm{dis}}[X_0 Y_1] = - 8 X_0 Y_1$, where $\mathcal{L}_{\mathrm{dis}}[O]$ is the second sum in (\ref{eq:1}). 
The explicit computation is easiest in the limit $c \gg 1$, where the Hamiltonian term in (\ref{eq:1}) is negligible compared to the dephasing term. In this limit (\ref{eq:16}) becomes (to leading order in $k$ and $c^{-1}$)
\begin{align}
    - i \Omega_{\bk} = \frac{k^2}{8 c} \sum_{A} (Z_0| \Lag^1 | A) (A| \Lag^1|Z_0) = - \frac{k^2}{c} \,,
\end{align}
where in the sum $A = X_0 Y_1$ and $Y_0 X_1$. The system is diffusive with $D = c^{-1} + O(c^{-2})$ for strong decoherence $c \gg 1$, cf. \cite{PhysRevE.92.042143}. This asymptotic behavior is verified numerically in Fig.\,\ref{fig1}, showing numerical results for finite $c$.

The numerical results are obtained using a truncated space of operators in (\ref{eq:7}) to evaluate (\ref{eq:16}). This is a different method compared to previous work, and is relatively straightforward to implement. It works best for larger values of $c$ where long operators are strongly suppressed by dissipation. These results on the XXZ example agree with those in the literature \cite{z,PhysRevLett.106.220601, men13, pro16, ljubotina2017spin}. In particular, for $0 < \Delta < 1$ the system is known to show ballistic spin transport in the absence of dephasing ($c=0$). Therefore, while transport is diffusive at nonzero dephasing, the diffusivity diverges as $c \to 0$.

{\it Constraints from the Lieb-Robinson bound.---} For diffusive rather than ballistic transport, $(C | \Lag^1 | C)$ must vanish in (\ref{eq:16}). Indeed, $|J) = \Lag^1 |C)$ is the current operator, and it is known from the Mazur-Suzuki bound \cite{MAZUR69, SUZUKI71} that if $(C | J) \neq 0$ at $k=0$, transport is ballistic. We restrict attention to non-ballistic systems \footnote{In the ballistic case, our argument gives a bound on the rate of attenuation of the linearly-dispersing mode}. Then the diffusivity can be rewritten as 
\be
D = - \left[ (C| \mathcal{L}^2 |C) + \int_0^\infty dt\, (C| \mathcal{L}^1 e^{\mathcal{L}^0 t} \mathcal{L}^1 |C) \right], \label{eq:23}
\ee
where $\frac{1}{E^0_a} |E^0_a)(E^0_a|$ in (\ref{eq:16}) has been replaced by an integral of $\exp \left(\Lag^0 t\right)$.
$D$ is manifestly real in (\ref{eq:23}) because in a basis of hermitian operators $\mathcal{L}^0,\mathcal{L}^2$ are real matrices and $\mathcal{L}^1$ is imaginary, and furthermore $(C|$ is a real vector.

To isolate the dynamics of the single conserved density we make a physical assumption about the spectrum of the Lindbladian decoherence operators: all local operators other than the charge density decay exponentially, at least as fast as $e^{- t / \tau}$. Here $\tau$ defines the `local decoherence time'.
The local difference in behavior between conserved densities and all other operators will be important for our argument. 
Technically, we will require
\begin{itemize}
\item {\bf Single-mode ansatz:} There exist $A, \tau > 0$ such that any local operator $O$ can be decomposed into local operators $O = \gamma I + O_1 + O_2$, where $I$ is the identity operator and $\gamma$ a coefficient, $O_1$ is a sum of $C$'s and $\|O_2(t)\| \leq A \|O\| e^{- t / \tau}$, $\|\dot{O}_2(t)\| \leq A \tau^{-1} \|O\| e^{- t / \tau}$ for $t >0$.
\end{itemize}
We will bound the diffusivity by the Lieb-Robinson velocity and the decoherence time $\tau$.

Let $\|\cdot\|$ be any operator norm contracted by the time evolution (\ref{eq:1}) \footnote{Completely positive unital maps --- such as Lindbladian time evolution --- between C*-algebras contract C*-norms, see Chap 8 of \cite{Nielsen} and Chap 3 of \cite{paulsen_2003}. }. This induces a seminorm (with $\||I)\| = 0$) for $|O) \in \mathcal{O}_0$:
\be\label{eq:induced}
    \||O)\| \equiv \lim_{\bk \to 0} \lim_{N \to \infty} N^{-1} \|\sum_{\bx} \Ts_{\bx}[O] e^{-i \bk \cdot \bx} \|,
\ee
where $N = \sum_{\bx} 1$ is the number of lattice sites.
For example, $\||Z)\| = \lim_{k \to 0} \lim_{N \to \infty} N^{-1} \|\sum_x Z_x e^{-i k x}\| = \|Z\|$,
and generally $\||O)\| \leq \|O\| \,,$ by the triangle inequality. From the definition (\ref{eq:induced}), this seminorm is also contracted by time evolution. As a result of contraction in time combined with the single-mode ansatz: 
\be
\||O)\| \geq \lim_{t \to \infty} \||O(t))\|  = |(C | O)| \||C)\|, \label{eq:22}
\ee 
bounding the norm of the $\bk=0$ state by its projection onto the conserved charge. 

We use (\ref{eq:22}) to bound the two terms in the diffusivity (\ref{eq:23}). For the first term, let $|O) = \mathcal{L}^2 |C) \in \mathcal{O}_0$. Then
\be
|(C | \mathcal{L}^2 | C)| \leq \| \mathcal{L}^2|C) \| / \||C)\| \,. \label{eq:28}
\ee
Given the operator equation of motion, the right-hand side of (\ref{eq:28}) is easily calculable.

To bound the second term in (\ref{eq:23}), take a local operator $O$ such that $|O) = \mathcal{L}^1 |C) + \alpha |C) \in \mathcal{O}_0$, with $\alpha \in \mathbb{C}$. Then:
\begin{align}\label{eq:odot}
    (C| \mathcal{L}^1 & e^{\mathcal{L}^0 t} \mathcal{L}^1 |C) = (C| \mathcal{L}^1 e^{\mathcal{L}^0 t} |O) \\
    & =   (C | \lim_{k\to 0} \partial_k (\mathcal{L} e^{\mathcal{L} t}) |O) =   (C | \lim_{k\to 0} \partial_k  | \dot{O}(t) ). \nonumber
\end{align}
The first equality uses $\mathcal{L}^0 |C) = 0$ and $(C | \mathcal{L}^1 | C) = 0$; the second equality uses $(C | \mathcal{L}^0 = 0$. 
In (\ref{eq:odot}), $\partial_k$ is defined to be the $k$-derivative of the components of matrices such as $\mathcal{L}$ or vectors such as $|O)$ in the prescribed basis $|O_a)$ in (\ref{eq:7}). Explicitly, for any local operator $P$ we can uniquely write $P = c I + \sum c^a_{\bx} \Ts_{\bx} [O_a]$ so that in $\mathcal{O}_{\bk}$,
$|P) = \sum c^a_{\bx} e^{i \bk \cdot \bx} |O_a)$ and  $- i \partial_k |P) = \sum c^a_{\bx}  (\hat{\bk} \cdot \bx) |\Ts_{\bx}[O_a])$, which is seen to be the first moment of the operator $P$.
Using (\ref{eq:odot}) in (\ref{eq:22}) gives the bound
\be
\left|\int_0^\infty dt\, (C| \mathcal{L}^1 e^{\mathcal{L}^0 t} \mathcal{L}^1 |C)\right| \leq \int_0^\infty d t \, \frac{\|\partial_k |\dot{O}(t))\|}{\||C)\|} \label{eq:29}.
\ee

Take an operator $J$ (the current) localized near the origin such that $|J) = \mathcal{L}^1 |C)$. According to the single-mode ansatz we can write $J = O + \sum_{\bx} c_{\bx} \Ts_{\bx}[C]$,  where $O$ is also localized near the origin and for $t > 0$
\be
     \|\dot{O}(t)\| \leq A \tau^{-1} \|J\| e^{- t / \tau}. \label{opdecay}
\ee
We can choose this $O$ as the operator in (\ref{eq:odot}).
From the bound (\ref{opdecay}) on $\|\dot{O}(t)\|$ we must now obtain a bound on $\|\partial_k |\dot{O}(t))\|$, that appears in (\ref{eq:29}).

Let $\mathcal{P}_l$ for $l \in \mathbb{R}$ be the projection onto the operator subspace spanned by all product operators  supported on the half-space $\hat{\bk} \cdot \bx \geq l$ and let $\mathcal{Q}_{l} = \text{Id} - \mathcal{P}_l$.
By an adaption \footnote{By \cite{commutator}, for any $l \in \mathbb{R}$ and $t > 0$ there exists $\tilde{O}$ localized in $\hat{\bk} \cdot \bx < l$ such that $\lVert\dot{O}(t) - \tilde{O}\rVert \leq A' \lVert\dot{O}\rVert e^{(v t - l) / \xi}$, hence $\lVert\mathcal{P}_{l}[\dot{O}(t)]\rVert = \lVert\mathcal{P}_{l}[\dot{O}(t) - \tilde{O}]\rVert \leq A' \lVert\dot{O}\rVert e^{(v t - l) / \xi}$. Note $\lVert\mathcal{P}_{l}[\dot{O}(t)]\rVert \leq \lVert\dot{O}(t)\rVert$ as well, hence $\lVert\mathcal{P}_{l}[\dot{O}(t)]\rVert \leq \min\{A' \lVert\dot{O}\rVert e^{(v t - l) / \xi}, \lVert\dot{O}(t)\rVert\}$ and plug in (\ref{opdecay}) to obtain (\ref{eq:30}). Similarly for (\ref{eq:31}) except that $\lVert\mathcal{Q}_{l}[\dot{O}(t)]\rVert = \lVert\dot{O}(t) - \mathcal{P}_{l}[\dot{O}(t)]| \leq 2 \lVert\dot{O}(t)\rVert$.} of the Lieb-Robinson bound \cite{PhysRevLett.104.190401, nachtergaele2011lieb} for (\ref{opdecay}), there exist $A' \geq 1$ and $v, \xi > 0$ such that for all $l \in \mathbb{R}, t > 0$, 
\begin{align}
\|\mathcal{P}_{l}[\dot{O}(t)]\| &\leq A \|J\|  \tau^{-1} \min\{e^{-t / \tau}, A' e^{(v t - l) / \xi} \}, \label{eq:30}\\
\|\mathcal{Q}_{l}[\dot{O}(t)]\| &\leq 2 A \|J\|  \tau^{-1} \min\{e^{-t / \tau}, A' e^{(v t + l) / \xi}\}. \label{eq:31}
\end{align}
The length $\xi$ is the range of microscopic interactions. We saw that $\partial_k$ corresponds to taking the first moment. Therefore $\partial_k|\dot{O}(t)) = i |O'(t))$, with
\begin{align}
O'(t) = \int_0^\infty d l\, \mathcal{P}_{l}[\dot{O}(t)] - \int_{-\infty}^0 d l\, \mathcal{Q}_{l}[\dot{O}(t)].
\end{align}
Indeed, from $\mathcal{P}_l[\mathcal{T}_{\bx}[O_a]]=\mathcal{T}_{\bx}[O_a]$ for $\hat{\bk} \cdot \bx \geq l$, and vanishing otherwise, we have $\int_0^\infty d l\, \mathcal{P}_l[\mathcal{T}_{\bx}[O_a]] = \hat{\bk} \cdot \bx\, \mathcal{T}_{\bx}[O_a]$ if $\hat{\bk} \cdot \bx \geq 0$, which is precisely the first moment. The second integral of $\mathcal{Q}_l$ similarly takes care of the $\hat{\bk} \cdot \bx \leq 0$ terms. Now, $\|\partial_k |\dot{O}(t))\| \leq \|O'(t)\|$ and, using (\ref{eq:30}) and (\ref{eq:31}), $\|O'(t)\| \leq 3 A \|J\| e^{-t / \tau} \tau^{-1} [v t +  \xi (1 + t / \tau + \ln A')]$. Hence, substituting into (\ref{eq:29}),
\be
\left|\int_0^\infty dt\, (C| \mathcal{L}^1 e^{\mathcal{L}^0 t} \mathcal{L}^1 |C)\right| \leq 3 A [v \tau + \xi (2 + \ln A')] \frac{\|J\|}{\||C)\|}. \nonumber
\ee

Putting the results together gives the diffusivity bound
\be\label{eq:fullbound}
D \leq D_0 + \left(\a \, v_\text{LR} \tau + \beta \, \xi \right) v_\text{C} \,.
\ee
Here $D_0 = \|\mathcal{L}^2|C)\| / \||C)\|$ is a `microscopic' diffusivity from the dissipative equation of motion. The Lieb-Robinson velocity $v_\text{LR} = v$ and $v_\text{C} = \|J\| / \||C)\|$ is a velocity obtained by dividing the current by the charge. As above $\tau$ is the decoherence time and $\xi$ is the range of microscopic interactions. The dimensionless coefficients $\a = 3 A$ and $\beta = 3 A (2 + \ln A')$.
Equation (\ref{eq:fullbound}) establishes that the diffusivity is bounded by microscopic velocities and time and lengthscales in the system.

The quantities $D_0$, $v_\text{LR}$, $v_\text{C}$ and $\xi$ can be obtained from the equations of motion. The quantities $A$ and $\tau$ are instead best determined experimentally or numerically from the decay of local non-conserved operators. From equation (\ref{opdecay}), $\tau$ determines the late time decay rate of the non-conserved part of the local current and $A = {\max_{t > 0} \| \dot{O}(t)\|/(\tau^{-1} \|J\| e^{-t/\tau})}$.
$A'$ does not have a strong effect as it appears in a logarithm in our bound.

{\it Final comments.---} The bound (\ref{eq:fullbound}) has nontrivial consequences for the dephasing XXZ chain. For $0 < \Delta < 1$ the diffusivity diverges in Fig.\,\ref{fig1} as $c \to 0$. The bound states that $D$ cannot diverge faster than $\tau$. 
In the XXZ model $v_\text{C} = 4$ and, from \cite{nachtergaele2011lieb}, $v_{\text{LR}} \leq 2 + \Delta$ are independent of $c$.
Now $\tau = { \max_k 1/(-\text{Re}\, E^1_k)}$, where $E^1_k \in \mathbb{C}$ is the first eigenvalue of $\partial_t|_{\ocal_k}$ above the slow mode. We evaluated this eigenvalue numerically by truncating the operator space as described around Fig.\,\ref{fig1}. At $\Delta = 0.5$ the ratio $D/\tau = 3.8(2)$ indeed remains finite as $c \to 0$.

We end with some broader comments. Firstly, (exponential) locality of interactions and a finite decoherence time are essential, as otherwise there can be superdiffusive transport \cite{PhysRevLett.108.093002, hild14, ljubotina2017spin}, where the perturbation theory (\ref{eq:16}) is no longer valid due to degeneracies or divergences. 

The decoherence-induced decay of operators such as long Pauli strings is phenomenologically similar to the decay of the thermal expectation values of those operators.
To obtain a rigorous bound on diffusion in unitary quantum dynamics in a thermal state, however, there will be several challenges to overcome. The diffusivity must be discussed in terms of expectation values rather than operators, and projections with respect to thermal inner products are difficult to evaluate (e.g. \S 5.6 of \cite{har18}). The butterfly velocity may causally constrain finite temperature transport \cite{Hartman:2017hhp, Lucas:2017ibu, Shenker:2013pqa, Roberts:2014isa, PhysRevLett.117.091602}, but a temperature-dependent bound on this velocity has not been established. These interesting problems are left for future work.

If a rigorous bound of the form $D \lesssim v^2 \tau + v \xi$ can indeed be established for diffusion in finite temperature states, it may shed light on the phenomenon of resistivity saturation \cite{RevModPhys.75.1085, hussey}. As temperature is increased $\tau$ will typically descrease, but $\xi$ is a microscopic and temperature-independent lengthscale. Therefore,  the resistivity $\rho \propto 1/D \gtrsim 1/(v^2 \tau + \xi v)$ is able to saturate at high temperatures where $v \tau < \xi$.

\section*{Acknowledgements}

The work of SAH is partially supported by DOE award DE-SC0018134. XH is supported by a Stanford Graduate Fellowship.

\bibliography{decoh}

\begin{thebibliography}{53}%
\makeatletter
\providecommand \@ifxundefined [1]{%
 \@ifx{#1\undefined}
}%
\providecommand \@ifnum [1]{%
 \ifnum #1\expandafter \@firstoftwo
 \else \expandafter \@secondoftwo
 \fi
}%
\providecommand \@ifx [1]{%
 \ifx #1\expandafter \@firstoftwo
 \else \expandafter \@secondoftwo
 \fi
}%
\providecommand \natexlab [1]{#1}%
\providecommand \enquote  [1]{``#1''}%
\providecommand \bibnamefont  [1]{#1}%
\providecommand \bibfnamefont [1]{#1}%
\providecommand \citenamefont [1]{#1}%
\providecommand \href@noop [0]{\@secondoftwo}%
\providecommand \href [0]{\begingroup \@sanitize@url \@href}%
\providecommand \@href[1]{\@@startlink{#1}\@@href}%
\providecommand \@@href[1]{\endgroup#1\@@endlink}%
\providecommand \@sanitize@url [0]{\catcode `\\12\catcode `\$12\catcode
  `\&12\catcode `\#12\catcode `\^12\catcode `\_12\catcode `\%12\relax}%
\providecommand \@@startlink[1]{}%
\providecommand \@@endlink[0]{}%
\providecommand \url  [0]{\begingroup\@sanitize@url \@url }%
\providecommand \@url [1]{\endgroup\@href {#1}{\urlprefix }}%
\providecommand \urlprefix  [0]{URL }%
\providecommand \Eprint [0]{\href }%
\providecommand \doibase [0]{http://dx.doi.org/}%
\providecommand \selectlanguage [0]{\@gobble}%
\providecommand \bibinfo  [0]{\@secondoftwo}%
\providecommand \bibfield  [0]{\@secondoftwo}%
\providecommand \translation [1]{[#1]}%
\providecommand \BibitemOpen [0]{}%
\providecommand \bibitemStop [0]{}%
\providecommand \bibitemNoStop [0]{.\EOS\space}%
\providecommand \EOS [0]{\spacefactor3000\relax}%
\providecommand \BibitemShut  [1]{\csname bibitem#1\endcsname}%
\let\auto@bib@innerbib\@empty
\bibitem [{\citenamefont {Gunnarsson}\ \emph {et~al.}(2003)\citenamefont
  {Gunnarsson}, \citenamefont {Calandra},\ and\ \citenamefont
  {Han}}]{RevModPhys.75.1085}%
  \BibitemOpen
  \bibfield  {author} {\bibinfo {author} {\bibfnamefont {O.}~\bibnamefont
  {Gunnarsson}}, \bibinfo {author} {\bibfnamefont {M.}~\bibnamefont
  {Calandra}}, \ and\ \bibinfo {author} {\bibfnamefont {J.~E.}\ \bibnamefont
  {Han}},\ }\href {\doibase 10.1103/RevModPhys.75.1085} {\bibfield  {journal}
  {\bibinfo  {journal} {Rev. Mod. Phys.}\ }\textbf {\bibinfo {volume} {75}},\
  \bibinfo {pages} {1085} (\bibinfo {year} {2003})}\BibitemShut {NoStop}%
\bibitem [{\citenamefont {Hussey}\ \emph {et~al.}(2004)\citenamefont {Hussey},
  \citenamefont {Takenaka},\ and\ \citenamefont {Takagi}}]{hussey}%
  \BibitemOpen
  \bibfield  {author} {\bibinfo {author} {\bibfnamefont {N.~E.}\ \bibnamefont
  {Hussey}}, \bibinfo {author} {\bibfnamefont {K.}~\bibnamefont {Takenaka}}, \
  and\ \bibinfo {author} {\bibfnamefont {H.}~\bibnamefont {Takagi}},\ }\href
  {\doibase 10.1080/14786430410001716944} {\bibfield  {journal} {\bibinfo
  {journal} {Philosophical Magazine}\ }\textbf {\bibinfo {volume} {84}},\
  \bibinfo {pages} {2847} (\bibinfo {year} {2004})}\BibitemShut {NoStop}%
\bibitem [{\citenamefont {Bruin}\ \emph {et~al.}(2013)\citenamefont {Bruin},
  \citenamefont {Sakai}, \citenamefont {Perry},\ and\ \citenamefont
  {Mackenzie}}]{Bruin804}%
  \BibitemOpen
  \bibfield  {author} {\bibinfo {author} {\bibfnamefont {J.~A.~N.}\
  \bibnamefont {Bruin}}, \bibinfo {author} {\bibfnamefont {H.}~\bibnamefont
  {Sakai}}, \bibinfo {author} {\bibfnamefont {R.~S.}\ \bibnamefont {Perry}}, \
  and\ \bibinfo {author} {\bibfnamefont {A.~P.}\ \bibnamefont {Mackenzie}},\
  }\href {\doibase 10.1126/science.1227612} {\bibfield  {journal} {\bibinfo
  {journal} {Science}\ }\textbf {\bibinfo {volume} {339}},\ \bibinfo {pages}
  {804} (\bibinfo {year} {2013})}\BibitemShut {NoStop}%
\bibitem [{\citenamefont {Schneider}\ \emph {et~al.}(2012)\citenamefont
  {Schneider}, \citenamefont {Hackerm{\"u}ller}, \citenamefont {Ronzheimer},
  \citenamefont {Will}, \citenamefont {Braun}, \citenamefont {Best},
  \citenamefont {Bloch}, \citenamefont {Demler}, \citenamefont {Mandt},
  \citenamefont {Rasch},\ and\ \citenamefont {Rosch}}]{schneider2012fermionic}%
  \BibitemOpen
  \bibfield  {author} {\bibinfo {author} {\bibfnamefont {U.}~\bibnamefont
  {Schneider}}, \bibinfo {author} {\bibfnamefont {L.}~\bibnamefont
  {Hackerm{\"u}ller}}, \bibinfo {author} {\bibfnamefont {J.~P.}\ \bibnamefont
  {Ronzheimer}}, \bibinfo {author} {\bibfnamefont {S.}~\bibnamefont {Will}},
  \bibinfo {author} {\bibfnamefont {S.}~\bibnamefont {Braun}}, \bibinfo
  {author} {\bibfnamefont {T.}~\bibnamefont {Best}}, \bibinfo {author}
  {\bibfnamefont {I.}~\bibnamefont {Bloch}}, \bibinfo {author} {\bibfnamefont
  {E.}~\bibnamefont {Demler}}, \bibinfo {author} {\bibfnamefont
  {S.}~\bibnamefont {Mandt}}, \bibinfo {author} {\bibfnamefont
  {D.}~\bibnamefont {Rasch}}, \ and\ \bibinfo {author} {\bibfnamefont
  {A.}~\bibnamefont {Rosch}},\ }\href {\doibase 10.1038/nphys2205} {\bibfield
  {journal} {\bibinfo  {journal} {Nature Physics}\ }\textbf {\bibinfo {volume}
  {8}},\ \bibinfo {pages} {213} (\bibinfo {year} {2012})},\ \bibinfo {note}
  {article}\BibitemShut {NoStop}%
\bibitem [{\citenamefont {Sagi}\ \emph {et~al.}(2012)\citenamefont {Sagi},
  \citenamefont {Brook}, \citenamefont {Almog},\ and\ \citenamefont
  {Davidson}}]{PhysRevLett.108.093002}%
  \BibitemOpen
  \bibfield  {author} {\bibinfo {author} {\bibfnamefont {Y.}~\bibnamefont
  {Sagi}}, \bibinfo {author} {\bibfnamefont {M.}~\bibnamefont {Brook}},
  \bibinfo {author} {\bibfnamefont {I.}~\bibnamefont {Almog}}, \ and\ \bibinfo
  {author} {\bibfnamefont {N.}~\bibnamefont {Davidson}},\ }\href {\doibase
  10.1103/PhysRevLett.108.093002} {\bibfield  {journal} {\bibinfo  {journal}
  {Phys. Rev. Lett.}\ }\textbf {\bibinfo {volume} {108}},\ \bibinfo {pages}
  {093002} (\bibinfo {year} {2012})}\BibitemShut {NoStop}%
\bibitem [{\citenamefont {Koschorreck}\ \emph {et~al.}(2013)\citenamefont
  {Koschorreck}, \citenamefont {Pertot}, \citenamefont {Vogt},\ and\
  \citenamefont {K{\"o}hl}}]{koschorreck2013universal}%
  \BibitemOpen
  \bibfield  {author} {\bibinfo {author} {\bibfnamefont {M.}~\bibnamefont
  {Koschorreck}}, \bibinfo {author} {\bibfnamefont {D.}~\bibnamefont {Pertot}},
  \bibinfo {author} {\bibfnamefont {E.}~\bibnamefont {Vogt}}, \ and\ \bibinfo
  {author} {\bibfnamefont {M.}~\bibnamefont {K{\"o}hl}},\ }\href {\doibase
  10.1038/nphys2637} {\bibfield  {journal} {\bibinfo  {journal} {Nature
  Physics}\ }\textbf {\bibinfo {volume} {9}},\ \bibinfo {pages} {405} (\bibinfo
  {year} {2013})}\BibitemShut {NoStop}%
\bibitem [{\citenamefont {Hild}\ \emph {et~al.}(2014)\citenamefont {Hild},
  \citenamefont {Fukuhara}, \citenamefont {Schau\ss{}}, \citenamefont {Zeiher},
  \citenamefont {Knap}, \citenamefont {Demler}, \citenamefont {Bloch},\ and\
  \citenamefont {Gross}}]{hild14}%
  \BibitemOpen
  \bibfield  {author} {\bibinfo {author} {\bibfnamefont {S.}~\bibnamefont
  {Hild}}, \bibinfo {author} {\bibfnamefont {T.}~\bibnamefont {Fukuhara}},
  \bibinfo {author} {\bibfnamefont {P.}~\bibnamefont {Schau\ss{}}}, \bibinfo
  {author} {\bibfnamefont {J.}~\bibnamefont {Zeiher}}, \bibinfo {author}
  {\bibfnamefont {M.}~\bibnamefont {Knap}}, \bibinfo {author} {\bibfnamefont
  {E.}~\bibnamefont {Demler}}, \bibinfo {author} {\bibfnamefont
  {I.}~\bibnamefont {Bloch}}, \ and\ \bibinfo {author} {\bibfnamefont
  {C.}~\bibnamefont {Gross}},\ }\href {\doibase 10.1103/PhysRevLett.113.147205}
  {\bibfield  {journal} {\bibinfo  {journal} {Phys. Rev. Lett.}\ }\textbf
  {\bibinfo {volume} {113}},\ \bibinfo {pages} {147205} (\bibinfo {year}
  {2014})}\BibitemShut {NoStop}%
\bibitem [{\citenamefont {Chien}\ \emph {et~al.}(2015)\citenamefont {Chien},
  \citenamefont {Peotta},\ and\ \citenamefont {Di~Ventra}}]{chien2015quantum}%
  \BibitemOpen
  \bibfield  {author} {\bibinfo {author} {\bibfnamefont {C.-C.}\ \bibnamefont
  {Chien}}, \bibinfo {author} {\bibfnamefont {S.}~\bibnamefont {Peotta}}, \
  and\ \bibinfo {author} {\bibfnamefont {M.}~\bibnamefont {Di~Ventra}},\ }\href
  {\doibase 10.1038/nphys3531} {\bibfield  {journal} {\bibinfo  {journal}
  {Nature Physics}\ }\textbf {\bibinfo {volume} {11}},\ \bibinfo {pages} {998}
  (\bibinfo {year} {2015})}\BibitemShut {NoStop}%
\bibitem [{\citenamefont {Luciuk}\ \emph {et~al.}(2017)\citenamefont {Luciuk},
  \citenamefont {Smale}, \citenamefont {B\"ottcher}, \citenamefont {Sharum},
  \citenamefont {Olsen}, \citenamefont {Trotzky}, \citenamefont {Enss},\ and\
  \citenamefont {Thywissen}}]{PhysRevLett.118.130405}%
  \BibitemOpen
  \bibfield  {author} {\bibinfo {author} {\bibfnamefont {C.}~\bibnamefont
  {Luciuk}}, \bibinfo {author} {\bibfnamefont {S.}~\bibnamefont {Smale}},
  \bibinfo {author} {\bibfnamefont {F.}~\bibnamefont {B\"ottcher}}, \bibinfo
  {author} {\bibfnamefont {H.}~\bibnamefont {Sharum}}, \bibinfo {author}
  {\bibfnamefont {B.~A.}\ \bibnamefont {Olsen}}, \bibinfo {author}
  {\bibfnamefont {S.}~\bibnamefont {Trotzky}}, \bibinfo {author} {\bibfnamefont
  {T.}~\bibnamefont {Enss}}, \ and\ \bibinfo {author} {\bibfnamefont {J.~H.}\
  \bibnamefont {Thywissen}},\ }\href {\doibase 10.1103/PhysRevLett.118.130405}
  {\bibfield  {journal} {\bibinfo  {journal} {Phys. Rev. Lett.}\ }\textbf
  {\bibinfo {volume} {118}},\ \bibinfo {pages} {130405} (\bibinfo {year}
  {2017})}\BibitemShut {NoStop}%
\bibitem [{\citenamefont {Wolf}\ \emph {et~al.}(2001)\citenamefont {Wolf},
  \citenamefont {Awschalom}, \citenamefont {Buhrman}, \citenamefont {Daughton},
  \citenamefont {von Moln{\'a}r}, \citenamefont {Roukes}, \citenamefont
  {Chtchelkanova},\ and\ \citenamefont {Treger}}]{Wolf1488}%
  \BibitemOpen
  \bibfield  {author} {\bibinfo {author} {\bibfnamefont {S.~A.}\ \bibnamefont
  {Wolf}}, \bibinfo {author} {\bibfnamefont {D.~D.}\ \bibnamefont {Awschalom}},
  \bibinfo {author} {\bibfnamefont {R.~A.}\ \bibnamefont {Buhrman}}, \bibinfo
  {author} {\bibfnamefont {J.~M.}\ \bibnamefont {Daughton}}, \bibinfo {author}
  {\bibfnamefont {S.}~\bibnamefont {von Moln{\'a}r}}, \bibinfo {author}
  {\bibfnamefont {M.~L.}\ \bibnamefont {Roukes}}, \bibinfo {author}
  {\bibfnamefont {A.~Y.}\ \bibnamefont {Chtchelkanova}}, \ and\ \bibinfo
  {author} {\bibfnamefont {D.~M.}\ \bibnamefont {Treger}},\ }\href {\doibase
  10.1126/science.1065389} {\bibfield  {journal} {\bibinfo  {journal}
  {Science}\ }\textbf {\bibinfo {volume} {294}},\ \bibinfo {pages} {1488}
  (\bibinfo {year} {2001})}\BibitemShut {NoStop}%
\bibitem [{\citenamefont {\ifmmode \check{Z}\else
  \v{Z}\fi{}uti\ifmmode~\acute{c}\else \'{c}\fi{}}\ \emph
  {et~al.}(2004)\citenamefont {\ifmmode \check{Z}\else
  \v{Z}\fi{}uti\ifmmode~\acute{c}\else \'{c}\fi{}}, \citenamefont {Fabian},\
  and\ \citenamefont {Das~Sarma}}]{RevModPhys.76.323}%
  \BibitemOpen
  \bibfield  {author} {\bibinfo {author} {\bibfnamefont {I.}~\bibnamefont
  {\ifmmode \check{Z}\else \v{Z}\fi{}uti\ifmmode~\acute{c}\else \'{c}\fi{}}},
  \bibinfo {author} {\bibfnamefont {J.}~\bibnamefont {Fabian}}, \ and\ \bibinfo
  {author} {\bibfnamefont {S.}~\bibnamefont {Das~Sarma}},\ }\href {\doibase
  10.1103/RevModPhys.76.323} {\bibfield  {journal} {\bibinfo  {journal} {Rev.
  Mod. Phys.}\ }\textbf {\bibinfo {volume} {76}},\ \bibinfo {pages} {323}
  (\bibinfo {year} {2004})}\BibitemShut {NoStop}%
\bibitem [{\citenamefont {Awschalom}\ and\ \citenamefont
  {Flatt{\'e}}(2007)}]{awschalom2007challenges}%
  \BibitemOpen
  \bibfield  {author} {\bibinfo {author} {\bibfnamefont {D.~D.}\ \bibnamefont
  {Awschalom}}\ and\ \bibinfo {author} {\bibfnamefont {M.~E.}\ \bibnamefont
  {Flatt{\'e}}},\ }\href {\doibase 10.1038/nphys551} {\bibfield  {journal}
  {\bibinfo  {journal} {Nature physics}\ }\textbf {\bibinfo {volume} {3}},\
  \bibinfo {pages} {153} (\bibinfo {year} {2007})}\BibitemShut {NoStop}%
\bibitem [{\citenamefont {Bogani}\ and\ \citenamefont
  {Wernsdorfer}(2008)}]{bogani2008molecular}%
  \BibitemOpen
  \bibfield  {author} {\bibinfo {author} {\bibfnamefont {L.}~\bibnamefont
  {Bogani}}\ and\ \bibinfo {author} {\bibfnamefont {W.}~\bibnamefont
  {Wernsdorfer}},\ }\href {\doibase 10.1038/nmat2133} {\bibfield  {journal}
  {\bibinfo  {journal} {Nature materials}\ }\textbf {\bibinfo {volume} {7}},\
  \bibinfo {pages} {179} (\bibinfo {year} {2008})}\BibitemShut {NoStop}%
\bibitem [{\citenamefont {Han}\ \emph {et~al.}(2014)\citenamefont {Han},
  \citenamefont {Kawakami}, \citenamefont {Gmitra},\ and\ \citenamefont
  {Fabian}}]{han2014graphene}%
  \BibitemOpen
  \bibfield  {author} {\bibinfo {author} {\bibfnamefont {W.}~\bibnamefont
  {Han}}, \bibinfo {author} {\bibfnamefont {R.~K.}\ \bibnamefont {Kawakami}},
  \bibinfo {author} {\bibfnamefont {M.}~\bibnamefont {Gmitra}}, \ and\ \bibinfo
  {author} {\bibfnamefont {J.}~\bibnamefont {Fabian}},\ }\href {\doibase
  10.1038/nnano.2014.214} {\bibfield  {journal} {\bibinfo  {journal} {Nature
  nanotechnology}\ }\textbf {\bibinfo {volume} {9}},\ \bibinfo {pages} {794}
  (\bibinfo {year} {2014})}\BibitemShut {NoStop}%
\bibitem [{\citenamefont {Ziman}(1960)}]{ziman}%
  \BibitemOpen
  \bibfield  {author} {\bibinfo {author} {\bibfnamefont {J.}~\bibnamefont
  {Ziman}},\ }\href {https://books.google.com/books?id=UtEy63pjngsC} {\emph
  {\bibinfo {title} {{Electrons and Phonons: The Theory of Transport Phenomena
  in Solids}}}},\ International series of monographs on physics\ (\bibinfo
  {publisher} {OUP Oxford},\ \bibinfo {year} {1960})\BibitemShut {NoStop}%
\bibitem [{\citenamefont {Schollwöck}(2011)}]{SCHOLLWOCK2011}%
  \BibitemOpen
  \bibfield  {author} {\bibinfo {author} {\bibfnamefont {U.}~\bibnamefont
  {Schollwöck}},\ }\href {\doibase https://doi.org/10.1016/j.aop.2010.09.012}
  {\bibfield  {journal} {\bibinfo  {journal} {Annals of Physics}\ }\textbf
  {\bibinfo {volume} {326}},\ \bibinfo {pages} {96 } (\bibinfo {year}
  {2011})},\ \bibinfo {note} {january 2011 Special Issue}\BibitemShut {NoStop}%
\bibitem [{\citenamefont {Karrasch}\ \emph {et~al.}(2012)\citenamefont
  {Karrasch}, \citenamefont {Bardarson},\ and\ \citenamefont {Moore}}]{kar12}%
  \BibitemOpen
  \bibfield  {author} {\bibinfo {author} {\bibfnamefont {C.}~\bibnamefont
  {Karrasch}}, \bibinfo {author} {\bibfnamefont {J.~H.}\ \bibnamefont
  {Bardarson}}, \ and\ \bibinfo {author} {\bibfnamefont {J.~E.}\ \bibnamefont
  {Moore}},\ }\href {\doibase 10.1103/PhysRevLett.108.227206} {\bibfield
  {journal} {\bibinfo  {journal} {Phys. Rev. Lett.}\ }\textbf {\bibinfo
  {volume} {108}},\ \bibinfo {pages} {227206} (\bibinfo {year}
  {2012})}\BibitemShut {NoStop}%
\bibitem [{\citenamefont {Leviatan}\ \emph {et~al.}(2017)\citenamefont
  {Leviatan}, \citenamefont {Pollmann}, \citenamefont {Bardarson},
  \citenamefont {Huse},\ and\ \citenamefont {Altman}}]{Leviatan:2017}%
  \BibitemOpen
  \bibfield  {author} {\bibinfo {author} {\bibfnamefont {E.}~\bibnamefont
  {Leviatan}}, \bibinfo {author} {\bibfnamefont {F.}~\bibnamefont {Pollmann}},
  \bibinfo {author} {\bibfnamefont {J.~H.}\ \bibnamefont {Bardarson}}, \bibinfo
  {author} {\bibfnamefont {D.~A.}\ \bibnamefont {Huse}}, \ and\ \bibinfo
  {author} {\bibfnamefont {E.}~\bibnamefont {Altman}},\ }\href@noop {} {\
  (\bibinfo {year} {2017})},\ \Eprint {http://arxiv.org/abs/1702.08894}
  {arXiv:1702.08894 [cond-mat.stat-mech]} \BibitemShut {NoStop}%
\bibitem [{\citenamefont {Bohrdt}\ \emph {et~al.}(2017)\citenamefont {Bohrdt},
  \citenamefont {Mendl}, \citenamefont {Endres},\ and\ \citenamefont
  {Knap}}]{knap17}%
  \BibitemOpen
  \bibfield  {author} {\bibinfo {author} {\bibfnamefont {A.}~\bibnamefont
  {Bohrdt}}, \bibinfo {author} {\bibfnamefont {C.~B.}\ \bibnamefont {Mendl}},
  \bibinfo {author} {\bibfnamefont {M.}~\bibnamefont {Endres}}, \ and\ \bibinfo
  {author} {\bibfnamefont {M.}~\bibnamefont {Knap}},\ }\href {\doibase
  10.1088/1367-2630/aa719b} {\bibfield  {journal} {\bibinfo  {journal} {New
  Journal of Physics}\ }\textbf {\bibinfo {volume} {19}},\ \bibinfo {pages}
  {063001} (\bibinfo {year} {2017})}\BibitemShut {NoStop}%
\bibitem [{\citenamefont {White}\ \emph {et~al.}(2018)\citenamefont {White},
  \citenamefont {Zaletel}, \citenamefont {Mong},\ and\ \citenamefont
  {Refael}}]{zal18}%
  \BibitemOpen
  \bibfield  {author} {\bibinfo {author} {\bibfnamefont {C.~D.}\ \bibnamefont
  {White}}, \bibinfo {author} {\bibfnamefont {M.}~\bibnamefont {Zaletel}},
  \bibinfo {author} {\bibfnamefont {R.~S.~K.}\ \bibnamefont {Mong}}, \ and\
  \bibinfo {author} {\bibfnamefont {G.}~\bibnamefont {Refael}},\ }\href
  {\doibase 10.1103/PhysRevB.97.035127} {\bibfield  {journal} {\bibinfo
  {journal} {Phys. Rev. B}\ }\textbf {\bibinfo {volume} {97}},\ \bibinfo
  {pages} {035127} (\bibinfo {year} {2018})}\BibitemShut {NoStop}%
\bibitem [{\citenamefont {Mazur}(1969)}]{MAZUR69}%
  \BibitemOpen
  \bibfield  {author} {\bibinfo {author} {\bibfnamefont {P.}~\bibnamefont
  {Mazur}},\ }\href {\doibase https://doi.org/10.1016/0031-8914(69)90185-2}
  {\bibfield  {journal} {\bibinfo  {journal} {Physica}\ }\textbf {\bibinfo
  {volume} {43}},\ \bibinfo {pages} {533 } (\bibinfo {year}
  {1969})}\BibitemShut {NoStop}%
\bibitem [{\citenamefont {Suzuki}(1971)}]{SUZUKI71}%
  \BibitemOpen
  \bibfield  {author} {\bibinfo {author} {\bibfnamefont {M.}~\bibnamefont
  {Suzuki}},\ }\href {\doibase https://doi.org/10.1016/0031-8914(71)90226-6}
  {\bibfield  {journal} {\bibinfo  {journal} {Physica}\ }\textbf {\bibinfo
  {volume} {51}},\ \bibinfo {pages} {277 } (\bibinfo {year}
  {1971})}\BibitemShut {NoStop}%
\bibitem [{\citenamefont {Prosen}(2011)}]{pro11}%
  \BibitemOpen
  \bibfield  {author} {\bibinfo {author} {\bibfnamefont {T.}~\bibnamefont
  {Prosen}},\ }\href {\doibase 10.1103/PhysRevLett.106.217206} {\bibfield
  {journal} {\bibinfo  {journal} {Phys. Rev. Lett.}\ }\textbf {\bibinfo
  {volume} {106}},\ \bibinfo {pages} {217206} (\bibinfo {year}
  {2011})}\BibitemShut {NoStop}%
\bibitem [{\citenamefont {Lieb}\ and\ \citenamefont
  {Robinson}(1972)}]{lieb1972}%
  \BibitemOpen
  \bibfield  {author} {\bibinfo {author} {\bibfnamefont {E.~H.}\ \bibnamefont
  {Lieb}}\ and\ \bibinfo {author} {\bibfnamefont {D.~W.}\ \bibnamefont
  {Robinson}},\ }\href {\doibase 10.1007/BF01645779} {\bibfield  {journal}
  {\bibinfo  {journal} {Comm. Math. Phys.}\ }\textbf {\bibinfo {volume} {28}},\
  \bibinfo {pages} {251} (\bibinfo {year} {1972})}\BibitemShut {NoStop}%
\bibitem [{\citenamefont {Poulin}(2010)}]{PhysRevLett.104.190401}%
  \BibitemOpen
  \bibfield  {author} {\bibinfo {author} {\bibfnamefont {D.}~\bibnamefont
  {Poulin}},\ }\href {\doibase 10.1103/PhysRevLett.104.190401} {\bibfield
  {journal} {\bibinfo  {journal} {Phys. Rev. Lett.}\ }\textbf {\bibinfo
  {volume} {104}},\ \bibinfo {pages} {190401} (\bibinfo {year}
  {2010})}\BibitemShut {NoStop}%
\bibitem [{\citenamefont {Hartnoll}(2015)}]{hartnoll15}%
  \BibitemOpen
  \bibfield  {author} {\bibinfo {author} {\bibfnamefont {S.~A.}\ \bibnamefont
  {Hartnoll}},\ }\href {\doibase 10.1038/nphys3174} {\bibfield  {journal}
  {\bibinfo  {journal} {Nature Physics}\ }\textbf {\bibinfo {volume} {11}},\
  \bibinfo {pages} {54} (\bibinfo {year} {2015})}\BibitemShut {NoStop}%
\bibitem [{\citenamefont {Prosen}(2014)}]{pro14}%
  \BibitemOpen
  \bibfield  {author} {\bibinfo {author} {\bibfnamefont {T.}~\bibnamefont
  {Prosen}},\ }\href {\doibase 10.1103/PhysRevE.89.012142} {\bibfield
  {journal} {\bibinfo  {journal} {Phys. Rev. E}\ }\textbf {\bibinfo {volume}
  {89}},\ \bibinfo {pages} {012142} (\bibinfo {year} {2014})}\BibitemShut
  {NoStop}%
\bibitem [{\citenamefont {Medenjak}\ \emph {et~al.}(2017)\citenamefont
  {Medenjak}, \citenamefont {Karrasch},\ and\ \citenamefont
  {Prosen}}]{PhysRevLett.119.080602}%
  \BibitemOpen
  \bibfield  {author} {\bibinfo {author} {\bibfnamefont {M.}~\bibnamefont
  {Medenjak}}, \bibinfo {author} {\bibfnamefont {C.}~\bibnamefont {Karrasch}},
  \ and\ \bibinfo {author} {\bibfnamefont {T.}~\bibnamefont {Prosen}},\ }\href
  {\doibase 10.1103/PhysRevLett.119.080602} {\bibfield  {journal} {\bibinfo
  {journal} {Phys. Rev. Lett.}\ }\textbf {\bibinfo {volume} {119}},\ \bibinfo
  {pages} {080602} (\bibinfo {year} {2017})}\BibitemShut {NoStop}%
\bibitem [{\citenamefont {Hartman}\ \emph {et~al.}(2017)\citenamefont
  {Hartman}, \citenamefont {Hartnoll},\ and\ \citenamefont
  {Mahajan}}]{Hartman:2017hhp}%
  \BibitemOpen
  \bibfield  {author} {\bibinfo {author} {\bibfnamefont {T.}~\bibnamefont
  {Hartman}}, \bibinfo {author} {\bibfnamefont {S.~A.}\ \bibnamefont
  {Hartnoll}}, \ and\ \bibinfo {author} {\bibfnamefont {R.}~\bibnamefont
  {Mahajan}},\ }\href {\doibase 10.1103/PhysRevLett.119.141601} {\bibfield
  {journal} {\bibinfo  {journal} {Phys. Rev. Lett.}\ }\textbf {\bibinfo
  {volume} {119}},\ \bibinfo {pages} {141601} (\bibinfo {year} {2017})},\
  \Eprint {http://arxiv.org/abs/1706.00019} {arXiv:1706.00019 [hep-th]}
  \BibitemShut {NoStop}%
\bibitem [{\citenamefont {Lindblad}(1976)}]{lindblad1976}%
  \BibitemOpen
  \bibfield  {author} {\bibinfo {author} {\bibfnamefont {G.}~\bibnamefont
  {Lindblad}},\ }\href {\doibase 10.1007/BF01608499} {\bibfield  {journal}
  {\bibinfo  {journal} {Comm. Math. Phys.}\ }\textbf {\bibinfo {volume} {48}},\
  \bibinfo {pages} {119} (\bibinfo {year} {1976})}\BibitemShut {NoStop}%
\bibitem [{\citenamefont {Breuer}\ and\ \citenamefont
  {Petruccione}(2007)}]{bre07}%
  \BibitemOpen
  \bibfield  {author} {\bibinfo {author} {\bibfnamefont {H.-P.}\ \bibnamefont
  {Breuer}}\ and\ \bibinfo {author} {\bibfnamefont {F.}~\bibnamefont
  {Petruccione}},\ }\href@noop {} {\emph {\bibinfo {title} {The Theory of Open
  Quantum Systems}}}\ (\bibinfo  {publisher} {Oxford University Press},\
  \bibinfo {year} {2007})\BibitemShut {NoStop}%
\bibitem [{\citenamefont {Nachtergaele}\ \emph {et~al.}(2011)\citenamefont
  {Nachtergaele}, \citenamefont {Vershynina},\ and\ \citenamefont
  {Zagrebnov}}]{nachtergaele2011lieb}%
  \BibitemOpen
  \bibfield  {author} {\bibinfo {author} {\bibfnamefont {B.}~\bibnamefont
  {Nachtergaele}}, \bibinfo {author} {\bibfnamefont {A.}~\bibnamefont
  {Vershynina}}, \ and\ \bibinfo {author} {\bibfnamefont {V.~A.}\ \bibnamefont
  {Zagrebnov}},\ }\href {\doibase 10.1090/conm/552} {\bibfield  {journal}
  {\bibinfo  {journal} {AMS Contemporary Mathematics}\ }\textbf {\bibinfo
  {volume} {552}},\ \bibinfo {pages} {161} (\bibinfo {year}
  {2011})}\BibitemShut {NoStop}%
\bibitem [{\citenamefont {Plenio}\ and\ \citenamefont
  {Knight}(1998)}]{RevModPhys.70.101}%
  \BibitemOpen
  \bibfield  {author} {\bibinfo {author} {\bibfnamefont {M.~B.}\ \bibnamefont
  {Plenio}}\ and\ \bibinfo {author} {\bibfnamefont {P.~L.}\ \bibnamefont
  {Knight}},\ }\href {\doibase 10.1103/RevModPhys.70.101} {\bibfield  {journal}
  {\bibinfo  {journal} {Rev. Mod. Phys.}\ }\textbf {\bibinfo {volume} {70}},\
  \bibinfo {pages} {101} (\bibinfo {year} {1998})}\BibitemShut {NoStop}%
\bibitem [{\citenamefont {\ifmmode \check{Z}\else
  \v{Z}\fi{}nidari\ifmmode~\check{c}\else \v{c}\fi{}}(2010)}]{z}%
  \BibitemOpen
  \bibfield  {author} {\bibinfo {author} {\bibfnamefont {M.}~\bibnamefont
  {\ifmmode \check{Z}\else \v{Z}\fi{}nidari\ifmmode~\check{c}\else
  \v{c}\fi{}}},\ }\href {\doibase 10.1088/1367-2630/12/4/043001} {\bibfield
  {journal} {\bibinfo  {journal} {New Journal of Physics}\ }\textbf {\bibinfo
  {volume} {12}},\ \bibinfo {pages} {043001} (\bibinfo {year}
  {2010})}\BibitemShut {NoStop}%
\bibitem [{\citenamefont {\ifmmode \check{Z}\else
  \v{Z}\fi{}nidari\ifmmode~\check{c}\else
  \v{c}\fi{}}(2011)}]{PhysRevLett.106.220601}%
  \BibitemOpen
  \bibfield  {author} {\bibinfo {author} {\bibfnamefont {M.}~\bibnamefont
  {\ifmmode \check{Z}\else \v{Z}\fi{}nidari\ifmmode~\check{c}\else
  \v{c}\fi{}}},\ }\href {\doibase 10.1103/PhysRevLett.106.220601} {\bibfield
  {journal} {\bibinfo  {journal} {Phys. Rev. Lett.}\ }\textbf {\bibinfo
  {volume} {106}},\ \bibinfo {pages} {220601} (\bibinfo {year}
  {2011})}\BibitemShut {NoStop}%
\bibitem [{\citenamefont {Mendoza-Arenas}\ \emph {et~al.}(2013)\citenamefont
  {Mendoza-Arenas}, \citenamefont {Al-Assam}, \citenamefont {Clark},\ and\
  \citenamefont {Jaksch}}]{men13}%
  \BibitemOpen
  \bibfield  {author} {\bibinfo {author} {\bibfnamefont {J.~J.}\ \bibnamefont
  {Mendoza-Arenas}}, \bibinfo {author} {\bibfnamefont {S.}~\bibnamefont
  {Al-Assam}}, \bibinfo {author} {\bibfnamefont {S.~R.}\ \bibnamefont {Clark}},
  \ and\ \bibinfo {author} {\bibfnamefont {D.}~\bibnamefont {Jaksch}},\ }\href
  {\doibase 10.1088/1742-5468/2013/07/P07007} {\bibfield  {journal} {\bibinfo
  {journal} {Journal of Statistical Mechanics: Theory and Experiment}\ }\textbf
  {\bibinfo {volume} {2013}},\ \bibinfo {pages} {P07007} (\bibinfo {year}
  {2013})}\BibitemShut {NoStop}%
\bibitem [{\citenamefont {Medvedyeva}\ \emph {et~al.}(2016)\citenamefont
  {Medvedyeva}, \citenamefont {Prosen},\ and\ \citenamefont {\ifmmode
  \check{Z}\else \v{Z}\fi{}nidari\ifmmode~\check{c}\else \v{c}\fi{}}}]{pro16}%
  \BibitemOpen
  \bibfield  {author} {\bibinfo {author} {\bibfnamefont {M.~V.}\ \bibnamefont
  {Medvedyeva}}, \bibinfo {author} {\bibfnamefont {T.}~\bibnamefont {Prosen}},
  \ and\ \bibinfo {author} {\bibfnamefont {M.}~\bibnamefont {\ifmmode
  \check{Z}\else \v{Z}\fi{}nidari\ifmmode~\check{c}\else \v{c}\fi{}}},\ }\href
  {\doibase 10.1103/PhysRevB.93.094205} {\bibfield  {journal} {\bibinfo
  {journal} {Phys. Rev. B}\ }\textbf {\bibinfo {volume} {93}},\ \bibinfo
  {pages} {094205} (\bibinfo {year} {2016})}\BibitemShut {NoStop}%
\bibitem [{\citenamefont {Ljubotina}\ \emph {et~al.}(2017)\citenamefont
  {Ljubotina}, \citenamefont {{\v{Z}}nidari{\v{c}}},\ and\ \citenamefont
  {Prosen}}]{ljubotina2017spin}%
  \BibitemOpen
  \bibfield  {author} {\bibinfo {author} {\bibfnamefont {M.}~\bibnamefont
  {Ljubotina}}, \bibinfo {author} {\bibfnamefont {M.}~\bibnamefont
  {{\v{Z}}nidari{\v{c}}}}, \ and\ \bibinfo {author} {\bibfnamefont
  {T.}~\bibnamefont {Prosen}},\ }\href {\doibase 10.1038/ncomms16117}
  {\bibfield  {journal} {\bibinfo  {journal} {Nature communications}\ }\textbf
  {\bibinfo {volume} {8}},\ \bibinfo {pages} {16117} (\bibinfo {year}
  {2017})}\BibitemShut {NoStop}%
\bibitem [{\citenamefont {Chaikin}\ and\ \citenamefont
  {Lubensky}(1995)}]{cha95}%
  \BibitemOpen
  \bibfield  {author} {\bibinfo {author} {\bibfnamefont {P.~M.}\ \bibnamefont
  {Chaikin}}\ and\ \bibinfo {author} {\bibfnamefont {T.~C.}\ \bibnamefont
  {Lubensky}},\ }\href@noop {} {\emph {\bibinfo {title} {Principles of
  Condensed Matter Physics}}}\ (\bibinfo  {publisher} {CUP},\ \bibinfo
  {address} {Cambridge},\ \bibinfo {year} {1995})\BibitemShut {NoStop}%
\bibitem [{Note1()}]{Note1}%
  \BibitemOpen
  \bibinfo {note} {Also, the identity operator $I$ itself doesn't contribute to
  ${\protect \mathcal {O}}_{\protect \boldsymbol {k}}$ for ${\protect
  \boldsymbol {k}}\not =0$ because then $(I)_{\protect \boldsymbol {k}} = 0$ in
  (\ref {eq:7})\label {foot:id}}\BibitemShut {NoStop}%
\bibitem [{\citenamefont {Forster}(1975)}]{for75}%
  \BibitemOpen
  \bibfield  {author} {\bibinfo {author} {\bibfnamefont {D.}~\bibnamefont
  {Forster}},\ }\href@noop {} {\emph {\bibinfo {title} {Hydrodynamic
  Fluctuations, Broken Symmetry and Correlation Functions}}}\ (\bibinfo
  {publisher} {Perseus Books},\ \bibinfo {year} {1975})\BibitemShut {NoStop}%
\bibitem [{\citenamefont {\ifmmode \check{Z}\else
  \v{Z}\fi{}nidari\ifmmode~\check{c}\else
  \v{c}\fi{}}(2015)}]{PhysRevE.92.042143}%
  \BibitemOpen
  \bibfield  {author} {\bibinfo {author} {\bibfnamefont {M.}~\bibnamefont
  {\ifmmode \check{Z}\else \v{Z}\fi{}nidari\ifmmode~\check{c}\else
  \v{c}\fi{}}},\ }\href {\doibase 10.1103/PhysRevE.92.042143} {\bibfield
  {journal} {\bibinfo  {journal} {Phys. Rev. E}\ }\textbf {\bibinfo {volume}
  {92}},\ \bibinfo {pages} {042143} (\bibinfo {year} {2015})}\BibitemShut
  {NoStop}%
\bibitem [{Note2()}]{Note2}%
  \BibitemOpen
  \bibinfo {note} {In the ballistic case, our argument gives a bound on the
  rate of attenuation of the linearly-dispersing mode}\BibitemShut {NoStop}%
\bibitem [{Note3()}]{Note3}%
  \BibitemOpen
  \bibinfo {note} {Completely positive unital maps --- such as Lindbladian time
  evolution --- between C*-algebras contract C*-norms, see Chap 8 of \cite
  {Nielsen} and Chap 3 of \cite {paulsen_2003}.}\BibitemShut {Stop}%
\bibitem [{Note4()}]{Note4}%
  \BibitemOpen
  \bibinfo {note} {By \cite {commutator}, for any $l \in \protect \mathbb {R}$
  and $t > 0$ there exists $\protect \mathaccentV {tilde}07E{O}$ localized in
  $\protect \mathaccentV {hat}05E{\protect \boldsymbol {k}} \cdot \protect
  \boldsymbol {x}< l$ such that $\delimiter 69645069 \protect \mathaccentV
  {dot}05F{O}(t) - \protect \mathaccentV {tilde}07E{O}\delimiter 86422285 \leq
  A' \delimiter 69645069 \protect \mathaccentV {dot}05F{O}\delimiter 86422285
  e^{(v t - l) / \xi }$, hence $\delimiter 69645069 \protect \mathcal
  {P}_{l}[\protect \mathaccentV {dot}05F{O}(t)]\delimiter 86422285 = \delimiter
  69645069 \protect \mathcal {P}_{l}[\protect \mathaccentV {dot}05F{O}(t) -
  \protect \mathaccentV {tilde}07E{O}]\delimiter 86422285 \leq A' \delimiter
  69645069 \protect \mathaccentV {dot}05F{O}\delimiter 86422285 e^{(v t - l) /
  \xi }$. Note $\delimiter 69645069 \protect \mathcal {P}_{l}[\protect
  \mathaccentV {dot}05F{O}(t)]\delimiter 86422285 \leq \delimiter 69645069
  \protect \mathaccentV {dot}05F{O}(t)\delimiter 86422285 $ as well, hence
  $\delimiter 69645069 \protect \mathcal {P}_{l}[\protect \mathaccentV
  {dot}05F{O}(t)]\delimiter 86422285 \leq \protect \qopname \relax
  m{min}\protect \{A' \delimiter 69645069 \protect \mathaccentV
  {dot}05F{O}\delimiter 86422285 e^{(v t - l) / \xi }, \delimiter 69645069
  \protect \mathaccentV {dot}05F{O}(t)\delimiter 86422285 \protect \}$ and plug
  in (\ref {opdecay}) to obtain (\ref {eq:30}). Similarly for (\ref {eq:31})
  except that $\delimiter 69645069 \protect \mathcal {Q}_{l}[\protect
  \mathaccentV {dot}05F{O}(t)]\delimiter 86422285 = \delimiter 69645069
  \protect \mathaccentV {dot}05F{O}(t) - \protect \mathcal {P}_{l}[\protect
  \mathaccentV {dot}05F{O}(t)]| \leq 2 \delimiter 69645069 \protect
  \mathaccentV {dot}05F{O}(t)\delimiter 86422285 $.}\BibitemShut {Stop}%
\bibitem [{\citenamefont {Hartnoll}\ \emph {et~al.}(2018)\citenamefont
  {Hartnoll}, \citenamefont {Lucas},\ and\ \citenamefont {Sachdev}}]{har18}%
  \BibitemOpen
  \bibfield  {author} {\bibinfo {author} {\bibfnamefont {S.~A.}\ \bibnamefont
  {Hartnoll}}, \bibinfo {author} {\bibfnamefont {A.}~\bibnamefont {Lucas}}, \
  and\ \bibinfo {author} {\bibfnamefont {S.}~\bibnamefont {Sachdev}},\
  }\href@noop {} {\emph {\bibinfo {title} {Holographic Quantum Matter}}}\
  (\bibinfo  {publisher} {MIT Press},\ \bibinfo {year} {2018})\BibitemShut
  {NoStop}%
\bibitem [{\citenamefont {Lucas}(2017)}]{Lucas:2017ibu}%
  \BibitemOpen
  \bibfield  {author} {\bibinfo {author} {\bibfnamefont {A.}~\bibnamefont
  {Lucas}},\ }\href@noop {} {\  (\bibinfo {year} {2017})},\ \Eprint
  {http://arxiv.org/abs/1710.01005} {arXiv:1710.01005 [hep-th]} \BibitemShut
  {NoStop}%
\bibitem [{\citenamefont {Shenker}\ and\ \citenamefont
  {Stanford}(2014)}]{Shenker:2013pqa}%
  \BibitemOpen
  \bibfield  {author} {\bibinfo {author} {\bibfnamefont {S.~H.}\ \bibnamefont
  {Shenker}}\ and\ \bibinfo {author} {\bibfnamefont {D.}~\bibnamefont
  {Stanford}},\ }\href {\doibase 10.1007/JHEP03(2014)067} {\bibfield  {journal}
  {\bibinfo  {journal} {JHEP}\ }\textbf {\bibinfo {volume} {03}},\ \bibinfo
  {pages} {067} (\bibinfo {year} {2014})},\ \Eprint
  {http://arxiv.org/abs/1306.0622} {arXiv:1306.0622 [hep-th]} \BibitemShut
  {NoStop}%
\bibitem [{\citenamefont {Roberts}\ \emph {et~al.}(2015)\citenamefont
  {Roberts}, \citenamefont {Stanford},\ and\ \citenamefont
  {Susskind}}]{Roberts:2014isa}%
  \BibitemOpen
  \bibfield  {author} {\bibinfo {author} {\bibfnamefont {D.~A.}\ \bibnamefont
  {Roberts}}, \bibinfo {author} {\bibfnamefont {D.}~\bibnamefont {Stanford}}, \
  and\ \bibinfo {author} {\bibfnamefont {L.}~\bibnamefont {Susskind}},\ }\href
  {\doibase 10.1007/JHEP03(2015)051} {\bibfield  {journal} {\bibinfo  {journal}
  {JHEP}\ }\textbf {\bibinfo {volume} {03}},\ \bibinfo {pages} {051} (\bibinfo
  {year} {2015})},\ \Eprint {http://arxiv.org/abs/1409.8180} {arXiv:1409.8180
  [hep-th]} \BibitemShut {NoStop}%
\bibitem [{\citenamefont {Roberts}\ and\ \citenamefont
  {Swingle}(2016)}]{PhysRevLett.117.091602}%
  \BibitemOpen
  \bibfield  {author} {\bibinfo {author} {\bibfnamefont {D.~A.}\ \bibnamefont
  {Roberts}}\ and\ \bibinfo {author} {\bibfnamefont {B.}~\bibnamefont
  {Swingle}},\ }\href {\doibase 10.1103/PhysRevLett.117.091602} {\bibfield
  {journal} {\bibinfo  {journal} {Phys. Rev. Lett.}\ }\textbf {\bibinfo
  {volume} {117}},\ \bibinfo {pages} {091602} (\bibinfo {year}
  {2016})}\BibitemShut {NoStop}%
\bibitem [{\citenamefont {Nielsen}\ and\ \citenamefont
  {Chuang}(2011)}]{Nielsen}%
  \BibitemOpen
  \bibfield  {author} {\bibinfo {author} {\bibfnamefont {M.~A.}\ \bibnamefont
  {Nielsen}}\ and\ \bibinfo {author} {\bibfnamefont {I.~L.}\ \bibnamefont
  {Chuang}},\ }\href@noop {} {\emph {\bibinfo {title} {Quantum Computation and
  Quantum Information: 10th Anniversary Edition}}},\ \bibinfo {edition} {10th}\
  ed.\ (\bibinfo  {publisher} {CUP},\ \bibinfo {address} {New York, NY, USA},\
  \bibinfo {year} {2011})\BibitemShut {NoStop}%
\bibitem [{\citenamefont {Paulsen}(2003)}]{paulsen_2003}%
  \BibitemOpen
  \bibfield  {author} {\bibinfo {author} {\bibfnamefont {V.}~\bibnamefont
  {Paulsen}},\ }\href {\doibase 10.1017/CBO9780511546631} {\emph {\bibinfo
  {title} {Completely Bounded Maps and Operator Algebras}}},\ Cambridge Studies
  in Advanced Mathematics\ (\bibinfo  {publisher} {CUP},\ \bibinfo {year}
  {2003})\BibitemShut {NoStop}%
\bibitem [{\citenamefont {Nachtergaele}\ \emph {et~al.}(2013)\citenamefont
  {Nachtergaele}, \citenamefont {Scholz},\ and\ \citenamefont
  {Werner}}]{commutator}%
  \BibitemOpen
  \bibfield  {author} {\bibinfo {author} {\bibfnamefont {B.}~\bibnamefont
  {Nachtergaele}}, \bibinfo {author} {\bibfnamefont {V.~B.}\ \bibnamefont
  {Scholz}}, \ and\ \bibinfo {author} {\bibfnamefont {R.~F.}\ \bibnamefont
  {Werner}},\ }in\ \href@noop {} {\emph {\bibinfo {booktitle} {Operator Methods
  in Mathematical Physics}}},\ \bibinfo {editor} {edited by\ \bibinfo {editor}
  {\bibfnamefont {J.}~\bibnamefont {Janas}}, \bibinfo {editor} {\bibfnamefont
  {P.}~\bibnamefont {Kurasov}}, \bibinfo {editor} {\bibfnamefont
  {A.}~\bibnamefont {Laptev}}, \ and\ \bibinfo {editor} {\bibfnamefont
  {S.}~\bibnamefont {Naboko}}}\ (\bibinfo  {publisher} {Springer Basel},\
  \bibinfo {address} {Basel},\ \bibinfo {year} {2013})\ pp.\ \bibinfo {pages}
  {143--149}\BibitemShut {NoStop}%
\end{thebibliography}%

\end{document}